\documentclass[prd,twocolumn,showpacs,preprintnumbers]{revtex4}
\pdfoutput=1
\usepackage{amsfonts,amsmath,amssymb}
\usepackage{graphicx}
\usepackage{color}
\usepackage{enumitem}
\usepackage{natbib}
\usepackage[plainpages=false, colorlinks=true, anchorcolor=blue, linkcolor=blue, citecolor=blue, bookmarks=false]{hyperref}
\pdfoutput=1

\begin{document}

\newcommand{\apjl}{Astrophys. J. Lett.}
\newcommand{\apjs}{Astrophys. J. Suppl. Ser.}
\newcommand{\aap}{Astron. \& Astrophys.}
\newcommand{\aj}{Astron. J.}
\newcommand{\araa}{Ann. Rev. Astron. Astrophys. } 
\newcommand{\mnras}{Mon. Not. R. Astron. Soc.}
\newcommand{\apss} {Astrophys. and Space Science}
\newcommand{\jcap}{JCAP}
\newcommand{\pasj}{PASJ}
\newcommand{\pasa}{Pub. Astro. Soc. Aust.}

\title{Statistical Significance of  spectral lag transition  in GRB 160625B}

\author{Shalini \surname{Ganguly}$^1$} \altaffiliation{E-mail: gangulyshalini1@gmail.com}
\author{Shantanu  \surname{Desai}$^1$} \altaffiliation{E-mail: shntn05@gmail.com}

\affiliation{$^{1}$Department of Physics, Indian Institute of Technology, Hyderabad, Telangana-502285, India}

\begin{abstract}
Recently Wei et al~\cite{Wei}  have found evidence for a transition from positive time lags to negative time lags in the spectral lag data of GRB 160625B. They have fit these observed lags to a sum of  two components: an assumed functional form for intrinsic time lag due to astrophysical mechanisms and an energy-dependent speed of light due to quadratic and linear Lorentz invariance violation (LIV) models. Here, we examine the statistical significance of the evidence for a transition to negative time lags. Such a transition, even if present in GRB 160625B, cannot be due to an energy dependent speed of light as this would contradict previous limits by some 3-4 orders of magnitude, and must therefore be of intrinsic astrophysical origin. We use three different model comparison techniques: a frequentist test and  two information based criteria (AIC and BIC). From the frequentist model comparison test, we find that the  evidence for transition in the spectral lag data is favored at $3.05\sigma$  and $3.74\sigma$  for the linear and quadratic models respectively. We find that $\Delta$AIC and $\Delta$BIC have values $\gtrsim$ 10 for the spectral lag transition that was motivated as being due to quadratic Lorentz invariance violating model pointing to ``decisive evidence''.  We note however that  none of the three models (including the model of intrinsic astrophysical emission)  provide a good fit to the data.

\end{abstract}

\maketitle

\section{Introduction}

In special relativity, the speed of light, $c$, is constant and has the same value in all inertial frames of reference. However, this 
\emph{ansatz} is no longer  true in Lorentz violating standard-model  extensions~\cite{Tasson} and also several quantum gravity and string theory models (see ~\cite{Mattingly,GAA} for reviews). In these models, Lorentz invariance is expected to be broken at very high energies close to the Planck scale, and 
the speed of light is dependent on the energy of the associated photon~\cite{GAA98}.
Although many astrophysical sources such as AGNs~\cite{AGN,Ellis13}, pulsars~\cite{Kaaret} etc. have been used to search for LIV-induced light speed variation, most of these searches have been done with Gamma-Ray Bursts (GRBs)(See ~\cite{Ellis,Abdo,Chang,Vasileiou13,Vasileiou15,Zhang,Xu1,Xu2} and references therein). 
Results from  searches for LIV prior to 2006 or so  can be found in the reviews in ~\cite{Mattingly,GAA}. We briefly enumerate some of the key results in the searches for this LIV since then.

Ellis et al~\cite{Ellis} considered a statistical sample of  about 60 GRBs at a range of redshifts  and modeled the  observed time-lag as sum of a constant intrinsic offset and an additional offset due to energy-dependent speed of light.  They found $4\sigma$  evidence that the  higher energy photons arrive earlier than the lower energy ones. The estimated lower limit was about $0.9 \times 10^{16}$ GeV. However, when an additional systematic offset was added to enforce the $\chi^2$/DOF for the null hypothesis to be of order unity, the statistical significance reduced to about $1\sigma$. 


Abdo et al~\cite{Abdo} have used the detection of multi-GeV photons  from a short GRB (GRB 090510), observed  within a one-second window by Fermi-LAT to obtain a robust limit on the LIV scale of greater than the Planck scale.  Vasileiou et al~\cite{Vasileiou13} applied three complementary techniques on four GRBs from Fermi-LAT, and the most stringent limit they obtain is from GRB090510 of about  7.6 times the Planck scale for a linear Lorentz invariance violation and 1.3 $\times 10^{11}$ GeV for quadratic Lorentz invariance violation. These limits also rule out results from ~\cite{Zhang,Xu1,Xu2}, who followed the same procedure  of \cite{Ellis} and claimed evidence for a linear correlation between the LIV induced time lag  and energy.

In contrast to the above searches, which looked for deterministic deviations in the speed of light as a function of energy, Vasileiou et al~\citep{Vasileiou15} looked for stochastic deviations in the speed of light, using high energy observations of GRB090510 from Fermi-LAT, and obtained a limit on the quantum gravity scale of more than twice the Planck scale at 95\% confidence level.

Most recently, Wei et al~\cite{Wei} (W17)  made an apparently  convincing case pertaining to the evidence for a transition from positive to negative time lag in the spectral lag data for GRB 160625B, by using the data from Fermi-LAT and Fermi-GBM. By modeling the time lag as sum of intrinsic time-lag (due to astrophysical processes) and energy-dependent speed of light due to Lorentz invariance violation (LIV),  which kicks in at high energies, they argued that this observation constitutes a robust evidence for a turnover in the spectral lag data. Subsequently, constraints on Lorentz invariance violation standard model extension coefficients have been obtained using this data~\cite{Wei2}. However, no quantitative assessment of the observed statistical significance  was made in these papers. 
In this work we compute the statistical significance  by using three  different model-comparison tests, namely frequentist hypothesis test, as well as information-criterion based tests.

The outline of this paper is as follows. We provide a succinct introduction to the model comparison  techniques used, in Section~\ref{sec:2}. 
We briefly review the observations, data analysis and conclusions reached by W17 in Section~\ref{sec:3}.  We then discuss the results from our model comparison tests  using the same data in Section~\ref{sec:4}. Our conclusions can be found in Section~\ref{sec:5}. 

\section{Introduction to Model Comparison Techniques}
\label{sec:2}

 In recent years a number of both Bayesian and frequentist model-comparison  techniques  (originally developed by the statistics community) have been applied to a variety of problems in astrophysics, cosmology, and particle physics to address  controversial issues. The aims of these techniques is two-fold. One is to find out which among the  two hypothesis is favored. A second goal is to assess the statistical significance or $p$-value of how well the better model is favored.
 We note however that in many of these applications, not all the techniques used reach the same conclusions. Also the  significances from the different techniques could be different. For our purpose, we shall employ multiple available techniques at our disposal to address how significant is the evidence for transition from positive to negative time lags in the spectral lag data. We briefly recap these techniques below. More details on each of these (from a  physics/astrophysics perspective) can be found in various reviews~\cite{Lyons,Liddle,Liddle07}. 
 
\begin{itemize}[leftmargin=*]

\item {\bf Frequentist Test}: The first step in a frequentist model  comparison test involves constructing a $\chi^2$ between a given  model and the data and then finding the best-fit parameters for each model. Then from the best-fit $\chi^2$ and degrees of freedom, one calculates the goodness of fit for each model, given by the $\chi^2$ probability or goodness of fit ~\citep{NR92}: 
\begin{equation}
P(\chi^2,\nu) = \dfrac{1}{2^{\frac{\nu}{2}}\Gamma(\nu/2)}(\chi^2)^{\frac{\nu}{2}-1}\exp\Big(-\frac{\chi^2}{2}\Big).
\label{eq:chiprob}
\end{equation}
where $\Gamma$ is the incomplete Gamma function and $\nu$ is the total degrees of freedom.

The best-fit model is the one with the larger value of $\chi^2$ goodness of fit. If the two models are nested, then from Wilk's theorem~\cite{Wilks}, the difference in $\chi^2$  between the two models satisfies a $\chi^2$ distribution with degrees of freedom equal to  the difference in the number of free parameters for the two hypotheses~\citep{Lyons}. Frequentist tests have been used a lot in astrophysics, from testing claims of sinusoidal variations in $G$ as a function of time~\cite{Desai16b}  to classification of GRBs~\cite{Kulkarni}.

\item {\bf Akaike Information Criterion}:
The Akaike Information Criterion (AIC) is used for model comparison, when we need to penalize for any additional free parameters  to avoid overfitting. AIC is an approximate  minimization of Kullback–-Leibler information entropy, which estimates the distance between two probability distributions~\citep{Liddle07}. For our purpose, we use the first-order corrected AIC, given by~\cite{Liddle}:
\begin{equation}
\rm{AIC} = \chi^2 + 2p  + \dfrac{2p(p+1)}{N-p-1},
\end{equation}
where $N$ is the total number of data points and  $p$ is the number of free parameters. A preferred model in this test is the one with the smaller value of AIC between the two hypothesis. From the difference in AIC ($\Delta$ AIC), there is no formal method to evaluate  a $p$-value~\footnote{See however~\cite{Shafer} which posits a significance based on $\exp(-\Delta AIC/2)$}. Only qualitative strength of evidence rules are available depending on the value of  $\Delta$AIC~\cite{Shi}.
\item {\bf Bayesian Information Criterion}:
The Bayesian Inference Criterion (BIC) is also used for penalizing the use of extra parameters. It is given by~\cite{Liddle}:
\begin{equation}
\rm{BIC} = \chi^2 + p \ln N .
\end{equation}
Similar to AIC, the model with the smaller value of BIC is the preferred model. 
The significance is estimated qualitatively in the same way as for AIC.  Both AIC and BIC have been used for comparison of cosmological models~\cite{Shi,Shafer,Liu}.

\end{itemize}

Besides these techniques, the ratio of Bayesian evidence (or odds ratio) ~\cite{Trotta} has also been extensively used for model comparison in astrophysics and particle physics~\cite{Shafer,Heavens,Trotta,Martin}. However, there have been criticisms regarding the usage of odds ratio for model comparison, since the Bayesian evidence depends on the priors chosen for the parameters~\cite{Cousins,Efth}. We shall not consider Bayesian evidence in this work.

\section{Summary of W17}
\label{sec:3}
W17 have used the spectral lag method to look for  energy-dependent time lags  in the arrival of photons of a  particular  GRB (namely GRB 160625B) using data from Fermi-LAT and Fermi-GBM, for which a remarkable transition from positive to negative time lags was observed in the arrival of higher energy photons.  The observation of photons from the same source is aimed at providing tighter constraints on Lorentz invariance violation factor. We now briefly describe the \emph{ansatz} made by W17 to fit the spectral lag data.

The observed time lags of photons of varying energies can be written down as :
\begin{equation}
\Delta t_{obs} = \Delta t_{int} + \Delta t_{LIV},
\label{eq:sum}
\end{equation}
where \(\Delta t_{int}\) is the intrinsic time lag between the emission of photon of a particular energy and the lowest energy photon from the GRB and $\Delta t_{LIV}$ is the time-lag due to Lorentz invariance violation (hereafter, LIV). The uncertainty associated with
$\Delta t_{int}$ is the largest, as it depends upon the internal dynamics of the GRB itself which cannot be obtained from observations. W17 posited the following model for  the intrinsic emission delay:
\begin{equation}
\Delta t_{int} (E) \rm{(sec)}  = \tau \left[\left(\frac{E}{keV}\right)^\alpha -  \left(\frac{E_0}{keV}\right)^\alpha \right] , 
\label{eq:int}
\end{equation}
\noindent where $E_0$=11.34 keV; whereas  $\tau$ and $\alpha$ are free parameters. This functional form was based on the observation  (from a recent study of the light curves of 50 GRBs), that most GRB light curves show positive time lags and the time tag is correlation with energy~\cite{Shao}. However, the analysis in ~\cite{Shao} was only up to energies of 400 keV, which is well below the possible transition energy (found by W17) of $\sim$ 8 MeV.
Therefore, there is no physics behind this particular functional form or evidence that this  function describes the spectral lag for all GRBs. Therefore, it has no advantage over other functional forms, which may provide a comparable or even a  better  fit to the data.
The  remaining time lag has  been attributed to the Lorentz violation effect, occurring  at a considerably higher energy  (closed to Planck scale)  and can be written as~\cite{Jacob}:
\begin{equation}
\Delta t_{LIV} = -\frac{1+n}{2H_0} \dfrac{E^n -E_0^n}{E^n_{QG,n}} \int_{0}^{z} \dfrac{(1+z^\prime)^n dz^\prime}{\sqrt{\Omega_M (1+z^\prime)^3 + \Omega_\Lambda}},
\label{eq:LIV}
\end{equation}
where $E_{QG,n}$ is the Lorentz-violating or quantum gravity scale, above which Lorentz violation kicks in; $H_0$ is the Hubble constant.
Studying the time lag of photons from a single source does not eliminate the necessity of taking into consideration the delay due to the intrinsic GRB mechanisms. Yet it does provide a more statistically robust method to fit the hypotheses to the observed data. Since the aim of this manuscript is to test the statistical significance of the claimed turnover in W17, we use the same parametric forms, namely  Eqn.~\ref{eq:int} and Eqn.~\ref{eq:LIV} to fit the spectral lag data.

GRB 160625B had three sub-bursts, where the time lag increased upto a certain photon energy after which it dramatically decreased. Using a fitting engine, {\tt McFit} (which uses Monte Carlo approach), W17 obtained the best fit parameters and their uncertainties corresponding to their proposed model consisting of the combined effects of intrinsic time lag and Lorentz violation.  As can be found in W17, the \(\chi^2_{dof}\) values are 2.39 and 2.25 for the linear (n=1) and quadratic (n=2) cases of LIV, respectively. These are however poor fits to the data, corresponding to $\chi^2$ probabilities of 
$3 \times 10^{-6}$ and $1.2 \times 10^{-5}$ respectively.
Using the best-fit values of \(\log E_{QG,1}\) and \(\log E_{QG,2}\) and their \(1 \sigma\) error bars, they obtained a \(1 \sigma\) lower limit on LIV as follows:
\begin{align*}
&E_{QG,1} \geqslant 0.5 \times 10^{16} \text{ GeV} \hspace{0.35cm} &( n = 1 )\\
&E_{QG,2} \geqslant 1.4 \times 10^{7} \text{ GeV} \hspace{0.35cm}	&( n = 2 )
\end{align*}

The main highlight of W17 was that they did not take into account a constant offset  for  intrinsic time lag as in Ellis et al~\cite{Ellis}. Instead, they proposed a power law function for the intrinsic time lag which fit well with the observed data.  They analyzed photons of different energy from the same source as compared to photons from several sources and simultaneously fit for the intrinsic and LIV-induced time lag; and their estimation of behavior of intrinsic time lag helped derive better limits of LIV.

Nevertheless, no estimate of the significance of transition in the spectral lag data compared to the null hypothesis of only astrophysically-induced time lag was done in W17. Also, all the models proved to be a bad fit to the data. We now independently reproduce the results and estimate the statistical significance of the turnover.

\section{Analysis}
\label{sec:4}
The first step in frequentist analysis involves parameter estimation for a given hypothesis by minimizing $\chi^2$ between the given model and data. We fit the data to  the same three hypotheses as in W17: the time lags are only due to the intrinsic astrophysical mechanisms given by Eqn.~\ref{eq:int}; followed by the hypothesis that the observed time lags consist of the sum of intrinsic  and LIV-induced time lags for linear (n=1 LIV)  as well as quadratic models (n=2 LIV).
Once we obtain the best-fit parameters, we then proceed to carry out model comparison using multiple techniques by  treating the only intrinsic emission case as the null hypothesis.

The best-fit values for the predicted models were obtained by  minimizing the $\chi^2$ functional~\cite{NR92} between the observed model and the data and using the observed errors in the time-lag as the errors in the ordinate.  We have assumed that the error bars between the different data points are uncorrelated. We also neglect the error bars in the X-axis.~\footnote{We also verified that our results do not change significantly by including the errors in $E$, which in this case correspond to the bin width, by modifying our $\chi^2$ to include errors in X-axis following the same prescription as in  \cite{Desai16b}.} In Eqn.~\ref{eq:LIV}, we used $H_0 = 67.3$ km/sec/Mpc and $\Omega_m$=0.315. These are same as those in W17 and inferred from Planck 2015 observations~\cite{Planck15}.

We fit the  37 spectral lag-energy measurements of GRB 160625B (data obtained from Table 1 of W17) to the three different hypotheses to obtain the  optimum values for \(\tau, \alpha\) and \(E_{qg}\). The best fit values obtained from $\chi^2$ minimization are summarized in Table 1. These mostly agree with the values obtained by W17~\footnote{We do not report 1$\sigma$ error bars, since our main goal is model comparison and not parameter estimation}.  The best-fit curves for all the three models along with the observed spectral lag data  are shown in Figure 1.
We see that for energies less than 15~MeV, $\Delta t_{obs}$ is correlated with energy. However above E $\sim$ 15.7 MeV, the observed time lag not only develops a negative correlation but it  becomes abruptly negative. The subsequent points after that again have positive values but display a gradual negative correlation with energy.

The best-fit values of $\chi^2$/ DOF for no LIV, LIV (n=1), LIV (n=2) are equal to 2.6, 2.37, and 2.23 respectively (cf. Table~\ref{tab:results}.) For a reasonably good fit, $\chi^2$/DOF has to be close to one~\cite{NR92}. Therefore, none of the models provide a decent fit to the data. The goodness of fit for each of these models is shown in Table~\ref{tab:results} and is less than $10^{-5}$. 

We  then proceed to carry out model comparison, by using the case of only intrinsic astrophysical emission (without a transition in the spectral lag data) as the null hypothesis. Among the three models, we compare the $\chi^2$ probability, P($\chi^2,\nu$) given by Eqn.~\ref{eq:chiprob}. From Table~\ref{tab:results}, we see that the model with the turnover due to  n=2 LIV  has the largest value of  P($\chi^2,\nu$) and hence
can be considered the best model amongst the three. In order to evaluate the statistical significance compared to the null hypothesis, we invoke Wilk's theorem, since the model of no LIV is nested within the  n=1 LIV and n=2 LIV models and can be recovered for $E_{QG}=\infty$. To evaluate the significance, we make use of the fact that the difference in $\chi^2$ between the  no LIV case
and the n=1 LIV and n=2 LIV models follows a $\chi^2$ distribution with degrees of freedom equal to one~\cite{Lyons}. From this, we calculate the $p$-value of n=1, by integrating the $\chi^2$ probability distribution from $\Delta \chi^2$
value between the two models to infinity. The $p$- values for n=1 LIV and n=2 LIV, compared to no LIV is equal to 
$0.0014$ and $9.2 \times 10^{-5}$ respectively. One way to interpret the $p$ value (for n=1), is that assuming the null hypothesis is true, the probability that we would see data that favors the model with n=1 LIV simply by chance is $0.0014$. We then define significance as the number of standard deviations that a Gaussian variable would fluctuate in one direction to give the same $p$-value~\cite{PDG}. We find that the significances of n=1 LIV and n=2 LIV correspond to $3.05\sigma$ and $3.74\sigma$ respectively. 

We  also obtained AIC and BIC difference values for these models as opposed to the null hypothesis (cf. Table~\ref{tab:results}). The model with the smaller AIC and BIC value is preferred but for our purpose of model comparison against the null hypothesis, we are mostly interested in the difference of AIC and BIC values. Both n=1 and n=2 LIV models have smaller AIC/BIC values compared to the null hypothesis. The $\Delta$AIC and $\Delta$BIC values for n=1 LIV  is about 8.5 and 6.9 respectively, which do not correspond to ``decisive evidence'',  according to the qualitative scales indicated in Shi et al~\cite{Shi}. For n=2 LIV, $\Delta$AIC and $\Delta$BIC correspond to  12.9 and 11.7 respectively, which denotes that the \emph{evidence against the null hypothesis is very strong}.


\begin{table}\label{tab}
\caption{Best-fit values of the models for the three hypotheses considered. The equation for the time lag with no LIV is  described in Eqn.~\ref{eq:int}. The equations for the two LIV models correspond to Eqn.~\ref{eq:sum}.}
\begin{center}
\begin{tabular}{|c |  c |   c |   c| }
\hline
&  \textbf{No LIV\color{blue}$^a$} & \textbf{ (n=1)\color{blue}$^b$} & \textbf{(n=2)\color{blue}$^c$}  \\
\hline
\textbf{$\alpha$}  &0.059 &0.175 &0.122 \\
\textbf{$\tau$ (sec)} &5.86 &1.24 &2.13 \\
\textbf{$E_{qg}/\rm{GeV}$} &  &4.7 $\times 10^{15}$ &1.47 $\times 10^7$ \\
\hline
\end{tabular}
\end{center}
\indent \indent $\color{blue}^a${\small \textit{No Lorentz Invariance}\\}
\indent \indent $\color{blue}^b${\small \textit{Lorentz Invariance up to linear (n=1) order}}\\
\indent\indent $\color{blue}^c${\small \textit{Lorentz Invariance up to quadratic (n=2) order}}\\
\end{table}

\includegraphics[scale=0.4]{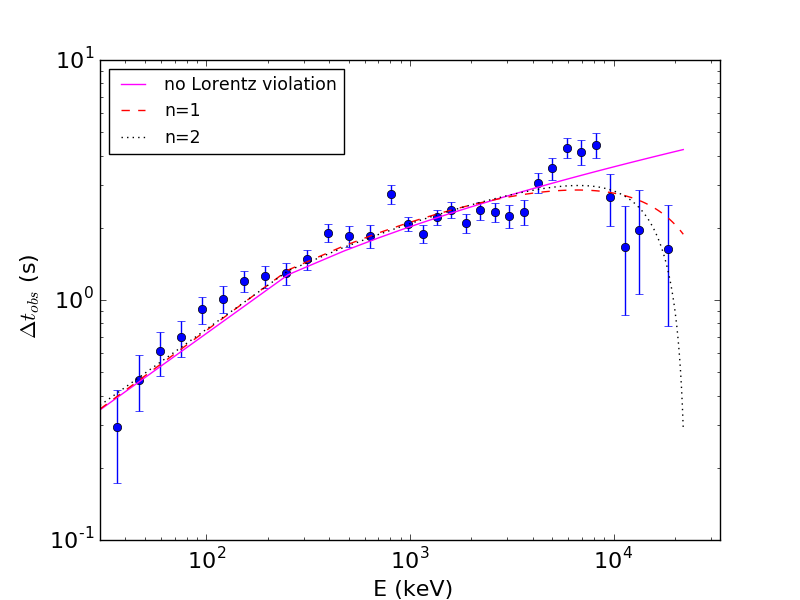}
\textbf{Figure 1 : }\small{Summary of the best fit LIV models for $n=1$ and $n=2$ along with  no Lorentz violation superposed on top of the  spectral lag data from GRB 160625B. We note that one data point at (E,$\Delta t$) = (15708 keV, -0.223 sec) has been omitted for brevity. All the spectral lag data points  have been obtained from Table 1 of W17.\\ }
\label{crab}



Although, all the three model comparison tests point to the  n=2 LIV case as the best-fit model, 
one possible concern  is that $\chi^2$/DOF is greater than two and the $\chi^2$ goodness of fit is less than about $10^{-5}$ for all the three models. 


\begin{table}
\caption{Statistical significance of Lorentz invariance violation (LIV)   for the two models (linear and quadratic LIV) as opposed to the null hypothesis, i.e. no Lorentz invariance violation using four different model comparison methods.  The frequentist  significance does not yet cross the $5\sigma$ threshold. Both $\Delta$AIC and $\Delta$BIC have values $>10$ for the quadratic LIV, pointing to ``decisive evidence'' using the qualitative strength of evidence rules.
However, all the three models have large values of $\chi^2$/DOF. So none of them (in an absolute sense) provide a good fit to the observed data.}
\label{tab:results}
\begin{center}
\begin{tabular}{|c| c | c | c|}
\hline
&  \textbf{No LIV\color{blue}$^a$} & \textbf{ (n=1)\color{blue}$^b$} & \textbf{(n=2)\color{blue}$^c$}  \\
\hline
\textbf{Frequentist}  & & & \\
DOF &35 &34 &34\\
\(\chi^2/\rm{DOF} \) & 2.6 &2.37 &2.23 \\
\(\chi^2 \rm{GOF} \) & 2.2$\times10^{-7}$ & 3.7$\times10^{-6}$ & 1.5$\times10^{-5}$ \\
$p$-value & &0.0014 &9.2$\times10^{-5}$ \\
significance & &3.05$\sigma$ &3.74$\sigma$ \\ \hline
\textbf{\(\Delta\) AIC}  &   &8.2   &12.9 \\
\textbf{\(\Delta\) BIC}  &  & 6.9 &11.7 \\ 
\hline
\end{tabular}
\end{center}
\indent \indent $\color{blue}^a${\small \textit{No Lorentz Invariance}\\}
\indent \indent $\color{blue}^b${\small \textit{Lorentz Invariance up to linear (n=1) order}}\\
\indent\indent $\color{blue}^c${\small \textit{Lorentz Invariance up to quadratic (n=2) order}}\\
\end{table}

\section{Conclusions}
\label{sec:5}
About a year ago Fermi-GBM and Fermi-LAT detected a remarkable Gamma-Ray Burst GRB160625B with three isolated sub-bursts  with a total duration of about 770 seconds. This GRB is the only known burst so far with a well-defined transition from positive to negative time lags
between photons of different energies. However, similar studies about the energy dependence of GRB intrinsic spectral lags over a wide energy range are very rare and have not been done systematically, making it is difficult to establish how common is such a transition from positive to negative spectral lags.
This spectral time-lag data was fit by W17~\cite{Wei} to a model consisting of an assumed functional form for intrinsic time lag caused by the astrophysical mechanism related to the GRB emission (see Eqn.~\ref{eq:int}) and a delay motivated as being due to energy-dependent speed of light, caused by the violation of Lorentz invariance. This Lorentz violation factor is a function of redshift and depends on  whether a linear or quadratic model is considered (see Eqn.~\ref{eq:LIV}). A joint fit was done to simultaneously determine the parameters of the intrinsic model and also the energy scale of Lorentz violation (or the quantum gravity scale). 
However, the $\chi^2$/DOF for both these models corresponds to 2.39 and 2.25  for linear and quadratic models. Both these models are therefore bad fits to the data 
with $\chi^2$ probabilities of  $3 \times 10^{-6}$ and $1.2 \times 10^{-5}$ respectively.
No statistical significance was estimated compared to the null hypothesis of only intrinsic astrophysical emission.

In this work, we redo the same analysis of the spectral lag data from GRB 160625B in order to estimate the statistical significance of the claim in W17. We fit the data to three different models. The first model posits that the time lag is only due to astrophysical emission and is considered the null hypothesis. The other two models involve a sum of the intrinsic mechanism and a linear  as well as  quadratic LIV model. The parameter estimation for all the three models was done by minimizing the $\chi^2$, similar to what was done in W17. We then carried out three different model comparison tests. The first test involves the frequentist comparison test, where we compare the $\chi^2$ probabilities, which is a proxy for the goodness of fit to determine the  model. 

Since the null hypothesis is nested within both the LIV models, we use Wilk's theorem to estimate the statistical significance of  the transition in the spectral lag data of  two LIV models compared to the null hypothesis of no violation of Lorentz invariance. We find that   the $\chi^2$/DOF for the null hypothesis, n=1 LIV, and  n=2 LIV  are equal to   2.6, 2.37, and 2.23  corresponding to $\chi^2$
probabilities of $2.2 \times 10^{-7}$, $3.7 \times 10^{-6}$, and $1.15 \times 10^{-5}$ respectively. Therefore, we find in agreement with the results of W17 that  all these models are a bad fit to the data. 
When we use Wilk's theorem and consider the case of no Lorentz violation as the null hypothesis, we find that the $p$-values are $1.4 \times 10^{-3}$ and $9.2 \times 10^{-5}$ corresponding to $3.05\sigma$ and
$3.74\sigma$ respectively. 
We also find that the
$\Delta$AIC and $\Delta$BIC are equal to 8.2 and 6.9 in favor of the n=1 LIV model compared to the null hypothesis. For n=2 LIV model, we find that  $\Delta \rm{AIC}$ and $\Delta$BIC are equal to 12.9 and 11.7. So the  information criterion based values just cross
the threshold for ``decisive evidence'' in favor of the n=2 LIV model.

Therefore, we conclude that the statistical significance of a fit to the turnover in the spectral lag data consisting of  an assumed functional form for intrinsic astrophysical emission and a second term motivated by Lorentz violation,  compared to the first term only (without any transition) is marginal. Even if such a transition exists in GRB 160625B, it cannot be of LIV origin as this violates previous limits by some 3-4 orders of magnitude (e.g.~\cite{Abdo,Vasileiou13}), and must therefore be of intrinsic origin. Improving our understanding of such intrinsic spectral lags at high energies could greatly help future LIV studies using GRBs.


\begin{acknowledgements}
We are grateful to Sabine Hossenfelder for drawing our attention to this result and also to the anonymous referee for invaluable feedback on our manuscript.
\end{acknowledgements}

\newpage

\end{document}